\documentclass[twocolumn,showpacs,preprintnumbers,amsmath,amssymb]{revtex4}
\usepackage{epsfig,amsfonts}
\usepackage{amsmath}
\usepackage{bbm}
\usepackage{graphicx,psfrag,rotating}
\usepackage{graphics}
\usepackage{txfonts}

\usepackage{graphicx}
\usepackage{dcolumn}
\usepackage{bm}

\newcommand{\ket}[1]{| #1\rangle}
\newcommand{\bra}[1]{\langle#1|}

\begin{document}


\title{Precision requirements for spin-echo based quantum memories}

\author{Khabat Heshami$^1$, Nicolas Sangouard$^2$, Ji\v{r}\'{i} Min\'{a}\v{r}$^2$, Hugues de Riedmatten$^{3,4}$ and Christoph Simon$^1$}
\affiliation{$^1$ Institute for Quantum Information Science
and Department of Physics and Astronomy, University of
Calgary, Calgary T2N 1N4, Alberta, Canada\\$^2$ Group of Applied Physics, University of Geneva, Switzerland \\$^3$ ICFO-Institute of Photonic Sciences,
Mediterranean Technology Park, 08860 Castelldefels
(Barcelona), Spain \\ $^4$ ICREA-Instituci\'o Catalana de Recerca i Estudis Avan\c cats, 08015 Barcelona, Spain}

\date{\today}

\begin{abstract}
Spin echo techniques are essential for achieving long
coherence times in solid state quantum memories for light
because of inhomogeneous broadening of the spin
transitions. It has been suggested that unrealistic levels
of precision for the radio frequency control pulses would
be necessary for successful decoherence control at the
quantum level. Here we study the effects of pulse
imperfections in detail, using both a semi-classical and a
fully quantum-mechanical approach. Our results show that
high efficiencies and low noise-to-signal ratios can be
achieved for the quantum memories in the single-photon
regime for realistic levels of control pulse precision. We also analyze errors due to imperfect initial state preparation (optical pumping), showing that they are likely to be more important than control pulse errors in many practical circumstances.
These results are crucial for future developments of solid
state quantum memories.
\end{abstract}

\maketitle
\section{\label{intro}Introduction }

Quantum memories for light \cite{lvovsky-natphotonics, simon-eur-reveiw} are key elements of quantum repeaters \cite{dlcz,sangouard09}, which are necessary to distribute entanglement over long distances for future quantum networks \cite{kimble08}. Quantum memories based on atomic ensembles \cite{lvovsky-natphotonics, simon-eur-reveiw, dlcz,eit,longdell05,raman,crib,hetet,hedges09,pre-afc,afc,hau09} are particularly attractive in practice because the light-matter coupling is enhanced by the large number of atoms and by collective interference effects. In the retrieval process collective interference can strongly enhance the re-emission of the stored light in a well-defined direction, compared to the non-directional background emission. This makes it possible to achieve high retrieval efficiencies \cite{hedges09,laurat,novikova,amari10} and small noise-to-signal ratios.

For long-distance applications such as quantum repeaters it
is essential for the memories to allow long storage times.
This can be achieved by using low-lying atomic states (spin
states) for storage \cite{lvovsky-natphotonics,
simon-eur-reveiw,dlcz,eit,longdell05,raman,afc,hau09}.
However, spin states are typically affected by
inhomogeneous broadening, i.e. different atoms in the
ensemble have slightly different energies. For atomic gases
this can be due to residual external magnetic fields or
intensity-dependent light shifts (for optical dipole
traps). In atomic gases it is possible to work with
field-insensitive clock transitions \cite{zhao09,
zhao-kumzich} to suppress inhomogeneous broadening due to
magnetic fields.

Solid-state atomic ensembles, such as rare-earth ion doped crystals, are attractive because there are no unwanted effects due to atomic motion and because solid-state systems promise enhanced scalability. However, they also have inhomogeneous broadening of the spin transitions. For example, in rare-earth doped crystals the rare-earth ions themselves produce a spatially varying potential due to spin-spin interactions \cite{fraval04,fraval-thesis}.

Inhomogeneous broadening is important because in the
absence of control techniques it limits the coherence time
of collective memory excitations to the inverse of the
inhomogeneous linewidth, which is typically in the tens of
microseconds range and is much shorter than the desired
storage times. This effect can be compensated using
spin-echo techniques, such as the application of a single
or a pair of $\pi$ pulses. The coherence time can be
further extended even beyond the single-atom $T_2$ time by
applying chains of $\pi$ pulses (bang-bang control)
\cite{longdell05, heinze10}.

In practice the control pulses are never perfect. In a recent experiment \cite{beavan} the most important imperfection was shown to be an inhomogeneity in the rf intensity across the sample, leading to a variation of about 1\% in the total pulse area seen by individual atoms.

In Ref. \cite{molmer04} the authors argued that for
successful operation in the quantum regime (i.e. when
single atomic excitations are stored) the $\pi$ pulses
would have to be precise to of order $1/N$, where $N$ is
the number of atoms. Typically, solid-state ensembles
contain of order $10^7$ to $10^9$ atoms, such a level of
precision would thus be completely out of reach. The
arguments of Ref. \cite{molmer04} were criticized in Ref.
\cite{longdell05}, but to our knowledge the question has
not been fully resolved until now.

In the present paper we study this problem in detail. The
argument of Ref. \cite{molmer04} was based on the fact that
imperfect $\pi$ pulses will lead to unwanted atomic
excitations, which will cause background emissions.
However, we will see that these emissions are
non-collective and hence non-directional. As a consequence,
good memory operation is achievable with realistic $\pi$
pulses.

There are several different ensemble-based quantum memory
protocols. For definiteness, in the following we focus on
the well-known Duan-Lukin-Cirac-Zoller (DLCZ) protocol
\cite{dlcz}. However, with small modifications our results
apply to many other protocols, including storage based on
electromagnetically induced transparency \cite{eit},
off-resonant Raman transitions \cite{raman}, controlled
reversible inhomogeneous broadening \cite{crib}, and atomic
frequency combs \cite{afc}.

\begin{figure}
\epsfig{file=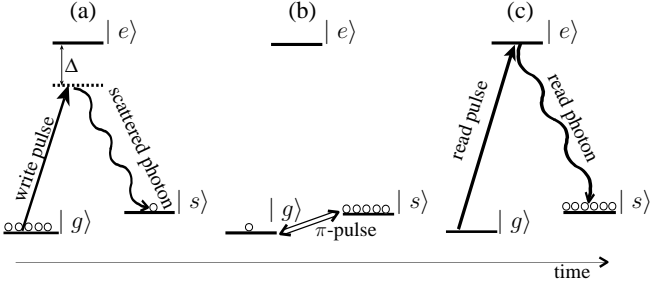,width=\linewidth}
\caption{\label{fig1} Basic level scheme in the Duan-Lukin-Cirac-Zoller protocol \cite{dlcz}. (a) The far detuned write pulse scatters a write photon and creates a single collective atomic excitation (spin-wave) in the state $s$. (b) Applying a $\pi$ pulse on the $g-s$ transition in the middle of the process interchanges the roles of $g$ and $s$, leading to rephasing at the end of storage period. (c) Shining the read pulse transforms the single collective atomic excitation in $g$ into a read photon.}
\end{figure}

In the DLCZ quantum memory protocol, as shown in Fig.\ref{fig1}(a), an off-resonant write pulse undergoes Raman scattering, leading to the creation of a single photon and a single collective excitation in the state $s$. This collective excitation dephases due to inhomogeneous broadening of the $g-s$ transition, but the application of a $\pi$ pulse in the middle of the storage time, see Fig.\ref{fig1}(b), can prepare a rephased atomic collective excitation at the time of retrieval. A read pulse can now be applied, which leads to the directional emission of the read photon, see Fig. \ref{fig1}(c).

We study the effects of spin-echo related imperfections in the described protocol. We begin with a semi-classical treatment for uniform errors in the $\pi$ pulses in section II. Then we give a fully quantum-mechanical treatment for the (most relevant) case of small $\pi$ pulse errors in section III, and we show that its results agree with the semi-classical approach. We treat the case of non-uniform $\pi$ pulse errors in appendix A. In section IV we consider the effects of imperfect optical pumping, i.e. an imperfect initial state. In section V we use the semi-classical approach to discuss the application of multiple $\pi$ pulses (bang-bang control). In section VI we give our conclusions.

\section{\label{semiclassical}Semi-classical approach}

We start with a semi-classical approach. Here the collective atomic state is treated as a tensor product of single-atom states. The single-excitation component of this tensor product corresponds to the true quantum state, which is why the two approaches give equivalent results in the single-photon regime, cf. section III below.

Consider an ensemble of $\Lambda$ type three-level atoms with two slightly split ground states $g$ and $s$, and an excited state $e$. At the beginning, all of the atoms are ideally pumped into the ground state $g$. Applying the write pulse that scatters a (Stokes) photon, transforms the state of the $k^{th}$ atom at the position $X_k$ into,
\begin{equation}\label{initialpsi}
\ket{\psi^{(k)}(t_0)}=\ket{g}-i\xi e^{i\vec{\Delta k}_{1}.\vec{X}_{k}} \ket{s},
\end{equation}
where $\xi$ represents the contribution of each atom to the collective excitation, and $\vec{\Delta k}_{1}=\vec{k}_w-\vec{k}_s$, where $\vec{k}_w$ is the $\vec{k}$-vector of the write pulse and $\vec{k}_s$ that of the scattered (Stokes) photon. For single-photon storage using an ensemble with $N$ atoms, $N\xi^2=1$. The total atomic state, as the product of single atomic states, implies that in the limit of $\xi\rightarrow 0$ the semi-classical atomic state tends to quantum mechanical state, see below. For weak light storage this limit is equivalent to limit of large number of atoms, $N\gg 1$.

Now, let us consider the effect of inhomogeneous broadening and the $\pi$ pulse. In the semi-classical picture one can easily use unitary evolution of a single atom in absence of the electromagnetic field to represent the dephasing (rephasing) before (after) applying the $\pi$ pulse. The propagator is given by,
\begin{eqnarray}\label{dephase}
U^{\Delta_{k}}(t_f,t_i)&=&\left(
\begin{array}{cc}
1 & 0\\
0 & e^{-i\Delta_{k}(t_f-t_i)}\\
\end{array}\right),
\end{eqnarray}
where $\Delta_k$ is the detuning from the central transition for the $k^{th}$ atom. The $\Delta_k$ have an inhomogeneous distribution with a width $\Gamma$. Then, the state of an atom after the time interval $\tau_{1}=t_1-t_0$ is $\ket{\psi^{(k)}(t_1)}=U^{\Delta_{k}}(t_1,t_0)\ket{\psi^{(k)}(t_0)}
$. An efficient retrieval is impossible for long storage times such that $\Gamma \tau \gtrsim 1$, since the readout amplitude is governed by the average atomic polarization, which is greatly reduced by the dephasing that takes place due to the temporal phase factors $e^{-i\Delta_{k}(t_f-t_i)}$ that vary from atom to atom. Applying the rf $\pi$ pulse, which is tuned to the central frequency of the $g-s$ transition, can bring this random phase into a negligible global phase at a certain time. In order to represent the $\pi$ pulse, let us recall the expression of the propagator of two levels of the atom under a pulsed excitation with the Rabi frequency $\Omega_i$ in the rotating wave approximation,
\begin{eqnarray}\label{atomresponse}
U^{\theta}(T)&=&\left(
\begin{array}{cc}
\cos(\theta_i/2) & -i\sin(\theta_i/2)\\
-i\sin(\theta_i/2) & \cos(\theta_i/2)\\
\end{array}\right),\\ \nonumber
\theta_i= \Omega_i T,
\end{eqnarray}
where $T$ is the temporal duration of the pulse.

The final state after applying a $\pi$ pulse at $t=t_1$ and
waiting the time interval $\tau_2=t_2-t_1$ is
$\ket{\psi^{(k)}(t_2)}=U^{\Delta_{k}}(t_2,t_1)U^{\theta}(T)U^{\Delta_{k}}(t_1,t_0)
\ket{\psi^{(k)}(t_0)}$, where $\theta=\pi\pm\epsilon$ and
$\theta=\pi$ represents a perfect $\pi$ pulse and
$\epsilon$ is the error. One can now retrieve the read
photon by applying the read pulse. Hence, we need to consider the spatial phase that comes from the read pulse to find the direction of the read photon. The spatial phase dependence due to the rf pulse can be ignored, because
$\vec{k}_{rf}.\vec{X}_{N} \ll 1$ for realistic sample
dimensions. Considering this fact, the final state after applying the read pulse is given by,
\begin{eqnarray}\label{psifinal}
\ket{\psi_f^{(k)}} &=&
e^{i\vec{k}_{r}.\vec{X}_k}(\cos(\theta/2)-\xi \sin(\theta/2)e^{-i\Delta_k\tau_1}e^{i\vec{\Delta k}_1.\vec{X}_k})\ket{e} \\ \nonumber
&-&ie^{-i\Delta_k \tau_2}
(\sin(\theta/2) + \xi e^{-i\Delta_k\tau_1} \cos(\theta/2) e^{i\vec{\Delta k}_1.\vec{X}_k})\ket{s},
\end{eqnarray}
where $\vec{k}_r$ is the $\vec{k}$-vector of the read pulse.

By transferring the population of the state $g$ into an excited state, the spin coherence transforms into an optical coherence, which leads to emission of the optical echo. The following shows how the atomic polarization would serve as the source of the echo signal,
\begin{eqnarray}\label{echointensity}
I_{echo}&=&I_0\frac{| P_{f}|^{2}}{\mu^2},\\ \nonumber
P_f &=& \sum_{k=1}^{N} \mu\langle e\mid \psi_f^{(k)}\rangle\langle \psi_f^{(k)}\mid s \rangle,
\end{eqnarray}
where, $\mu$ is the electric dipole moment and $I_0$ is the radiation intensity of one isolated atom. It can be seen from eq. (\ref{psifinal}) that terms with atom-dependent temporal phases appear in the atomic polarization in the eq. (\ref{echointensity}). Since the emission is governed by the average over single atomic polarizations, only those terms for which these phases are canceled contribute significantly to the echo intensity. Analyzing different terms in $\langle e\mid \psi_f^{(k)}\rangle \langle \psi_f^{(k)} \mid s\rangle$ one finds that the term $i\xi\sin^2(\theta/2)e^{i\Delta_k(\tau_2-\tau_1)}e^{i(\vec{\Delta k}_1+\vec{k}_{r}).\vec{X}_k}$ is the only one for which one can exclude the atom-dependent temporal phase, which is called rephasing. This can take place under the condition $\tau_1=\tau_2$, which implies that the $\pi$ pulse has to be applied at the middle of the process.

\begin{figure}
\epsfig{file=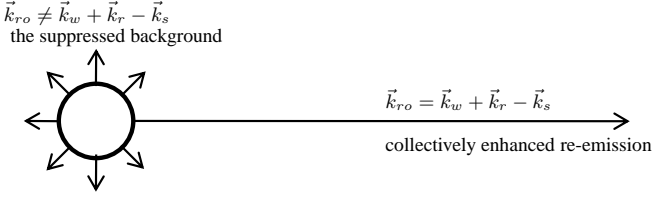,width=\linewidth}
\caption{\label{fig2} A schematic representation of the collective enhancement at a certain direction that is given by the phase-matching condition. Non-directional re-emission is suppressed by a factor $1/N$, where $N$ is the number of atoms.}
\end{figure}

The dipole moment of each single atom serves as a source of radiation. In the far field approximation and under the rephasing condition $(\tau_1=\tau_2)$, one can show that the amplitude of the readout from the whole ensemble in the direction $\vec{k}_{ro}$ is proportional to $\sum_{j=1}^{N}e^{i(\vec{\Delta k}_1+\vec{\Delta k}_2).\vec{X}_j} \sin^2{\theta/2}$, where $\vec{\Delta k}_2= \vec{k}_{r}-\vec{k}_{ro}$. Hence, for $\vec{\Delta
k}_1+\vec{\Delta k}_2=0$ the readout intensity is
proportional to $N^2$. This corresponds to constructive
interference of the re-emission from all of the atoms at a
certain direction $\vec{k}_{ro} = \vec{k}_{w} + \vec{k}_{r}
- \vec{k}_{s}$. However, even by applying an ideal $\pi$
pulse the radiation at other directions is not zero. The
background re-emission (non-directional) intensity is
proportional to $\langle \sum_{j,k=1}^{N} e^{i(\vec{\Delta
k}_1+\vec{\Delta k}_2).(\vec{X}_j-\vec{X}_k)}\rangle = N$, where $\vec{\Delta k}_1+\vec{\Delta k}_2 \neq 0$.
Accordingly, the intensity ratio of the collectively
enhanced re-emission (directional) and the randomly
distributed background re-emission (non-directional) is
$N$, see Fig. (\ref{fig2}). For example, for
counter-propagation of the write and read pulses the
phase-matching condition gives $\vec{k}_{ro}=-\vec{k}_{s}$.

However, the error in the $\pi$ pulse prevents achieving the highest possible the echo amplitude and presents a source of noise.

{\it Efficiency reduction.} By considering these points one can calculate the atomic polarization, $P_f =-iN\xi \mu \cos^{2}(\epsilon/2)$, because $\theta=\pi\pm \epsilon$. Consequently, the intensity of the echo is,
\begin{equation}\label{semiintecho}
I_{echo}=I_0 N^2 \xi^{2} \cos^{4}(\epsilon/2),
\end{equation}
where $I_0$ has the same definition as in eq. (\ref{echointensity}). Eventually, the total efficiency of any process depends on the optical depth, performance of the experimental facilities and other theoretical and experimental details which are related to the protocol and the experimental setup. However, the present result allows to find the efficiency reduction due to the error in the $\pi$ pulse that is important in our analysis.

{\it Noise-to-signal ratio.} Due to error in the rephasing pulse ($\epsilon \neq 0$) there is a noise in the re-emission. It is important to analyze the intensity of the noise, because for high memory fidelity the noise to echo ratio has to be small.

A single $\pi$ pulse interchanges the states of the atoms from $s$ to $g$ and vice versa. However, an error in the $\pi$ pulse produces population in the ground state $g$ even without applying write pulse at the beginning. This would lead to fluorescent radiation. Hence, it can be distinguished by the non-zero terms in $\ket{\psi_f^{(k)}}$ for $\xi=0$, which means even with no write pulse the error by itself can produce population in $g$. Quantitatively, the fluorescent radiation can be specified by $\mid \langle g \mid {\psi^{(k)}_{f}}^{\xi=0}\rangle\mid^2=\cos^2{(\theta/2)}$ which is the term in $\ket{\psi_{f}}$ that does not originate from the read-in process ($\xi=0$), but gives a non-zero projection on $\ket{g}$. Therefore, the intensity of the noise that originates from error in the $\pi$ pulse can be given by,
\begin{eqnarray}\label{seminoiseint}
I_{noise} &=& I_0 \sum_{k=1}^{N} \mid \langle g\mid {\psi^{(k)}_{f}}^{\xi=0}\rangle \mid^{2} \\ \nonumber
&=& I_0 N \sin^{2}(\epsilon/2).
\end{eqnarray}
Obviously, the fluorescent radiation as the source of the noise has the spatial dependence of a single photon radiation. Consequently, the semi-classical picture yields an equally distributed noise (non-directional) that comes from single atom radiation of ensemble of the atoms, because of error in the rephasing pulse.

In order to analyze the effect of an imperfect $\pi$ pulse on the fidelity, it is important to study the noise-to-signal ratio. The following presents the noise-to-signal ratio,
\begin{equation}\label{semiratio}
r=\frac{I_{noise}}{I_{echo}} =\frac{\sin^{2}(\epsilon/2)}{\cos^{4}(\epsilon/2)},
\end{equation}
keeping in mind that $N\xi^2=1$ for single-photon storage.
The higher the noise-to-signal ratio, the less fidelity we
have. In \cite{beavan}, 1\% variation has been realized in
the intensity of the rf pulse that causes the same error in
the pulse area. Such error gives quite low noise-to-signal
ratio of $0.25\times 10^{-4}$ and only wastes 0.1\% of the
efficiency. The semi-classical calculation suggests that
the typical 1\% error in the $\pi$ pulse which is far
beyond the $1/N$ precision, does not impose a major
constraint on the efficiency and fidelity. In addition to
the semi-classical calculations, it is interesting to
reconsider the problem using the quantum mechanical
description. In the next section, we perform fully quantum
mechanical investigation for the same question.

\section{\label{quantumapproach} Quantum mechanical treatment}
In this section, we perform fully quantum mechanical analysis on a global error in the $\pi$ pulse. We consider an ideal initial state where all the atoms have been pumped into the ground state.

\subsection{Ideal Protocol}

As we discussed before, the large detuning of the write laser from $g-e$ leads to scattering of a photon and creating a collective atomic excitation. The theory of light-atom interaction is well established to describe interaction between the field and the atomic dipole moment. In general, one can represent the interaction Hamiltonian as,
\begin{equation}\label{intham}
H_{int}=\sum_{j=1}^{N} G \int d{\vec k}  {\hat a}(\vec{k}) e^{i\vec{k}.\vec{X}_j} {\hat \sigma}_{\rho\nu}^{j}+h.c.,
\end{equation}
where $G=\langle \rho| {\hat \mu}_j.\vec{\epsilon}| \nu\rangle \sqrt{\frac{\hbar \omega}{2\epsilon_0 V}}$ and ${\hat \sigma}_{\rho\nu}^{j}=| \rho\rangle_{jj}\langle \nu|$, where $\rho$ and $\nu$ denote atomic levels $g$, $e$ and $s$. This Hamiltonian shows the interaction between a field and the dipole moment of the $j^{th}$ atom, ${\hat \mu}_j$. For simplicity, we assume the dipole-field coupling is identical for all the atoms and we have not considered the transverse profile. One can extract the interaction Hamiltonian for the write laser and the scattered photon from eq. (\ref{intham}), see FIG.\ref{fig1}(a). By combining the interaction Hamiltonians and considering adiabatic elimination of the excited, $e$, one can describe the read-in part of the process as follows,
\begin{equation}\label{Hint}
H_{int}^{eff}= \sum_{j=1}^{N} G^{\prime} \int d\vec{k}_s  {\hat a}^{\dagger}_{s}(\vec{k}_s) e^{i(\vec{k}_{w}-\vec{k}_{s}).\vec{X}_j} {\hat \sigma}_{sg}^{j}+h.c.,
\end{equation}
where $G^{\prime}=\langle s| {\hat \mu }_j.\vec{\epsilon}_{s}| e\rangle \langle e| {\hat \mu}_j.\vec{\epsilon}_{w}| g\rangle \sqrt{\frac{\hbar \omega_{s}}{2\epsilon_0 V}}\varepsilon_{w}(\tau)$ and $\Omega_{w}(\tau)=\langle e| {\hat \mu}_j.\vec{\epsilon}_{w}| g\rangle \varepsilon_{w}(\tau)$ is the Rabi frequency of the classical write field.

The unitary evolution under this effective interaction Hamiltonian describes the creation of a Stokes photon via Raman scattering, which is accompanied by the creation of a collective atomic state. The collective state is
\begin{eqnarray}\label{colexc}
\ket{\psi(t_0)} &=& \frac{1}{\sqrt{N}}\sum_{k=1}^{N}e^{i\vec{\Delta k}_1.\vec{X}_k}\ket{g...s^{(k)}...g},
\end{eqnarray}
where $\ket{g...s^{(k)}...g}$ shows all atoms in the ground
state and the $k^{th}$ atom in the spin state $s$ and
$\vec{\Delta k}_1=\vec{k}_w-\vec{k}_s$. Here, we neglect
the terms with more than one excitation, because we are
interested in single photon storage. The interaction
Hamiltonian governs the evolution of the atomic and the
photonic state of the system. Our analysis follows the
evolution of the atomic state. However, later in this
section, we will refer to the above discussion in order to
find the intensity of the readout based on the norm of the
final atomic state.

One can find the state $\ket{\psi(t_0)}$ in eq. (\ref{colexc}) by applying the Schwinger bosonic creation operator or equally weighted superposition of $\sigma_{+}=\mid s \rangle \langle g \mid$ operators, $J_{+}(\vec{\Delta k}_1)=\sum_{k=1}^{N} e^{i\vec{\Delta k}_1.\vec{X}_k} \sigma_{+}^{(k)}\otimes \mathbbm{1}$. The inhomogeneous spin broadening implies that each atom has a slightly different energy than the other atom's spin level that indicates any atom will evolve based on its small energy detuning specified by $\Delta_n$. After the time interval $\tau_1=t_1-t_0$ the state $\ket{\psi(t_0)}$ will be evolved to $\ket{\psi(t_1)} = \frac{1}{\sqrt{N}}\sum_{k=1}^{N} e^{i\Delta_k \tau_1}e^{i\vec{\Delta k}_1.\vec{X}_k}\ket{g...s^{(k)}...g}$. The effect of dephasing as a result of inhomogeneous spin broadening can be described as an atom-dependent phase accumulation of the single collective atomic excitation.

The Following operator is appropriate to mathematical modeling of the dephasing after the time $t$,
\begin{eqnarray}\label{phaope}
e^{i\hat{\Omega}t}=\otimes_{k=1}^{N} e^{i\Delta_{k}t\ket{s}_{k}\bra{s}}, \\ \nonumber
\end{eqnarray}
where
$\hat{\Omega}=\sum_{k=1}^{N}\Delta_{k}\ket{s}_{k}\bra{s}\otimes
\mathbbm{1}$, see \cite{footnote1}, where the operator
$\mathbbm{1}$ shows the identity operator that acts on the
rest of the atoms. The operator in Eq. (\ref{phaope}) can be used to show the
rephasing after applying the $\pi$ pulse. Hence, one can
represent the dephased state as
$\ket{\psi(t_1)}=\frac{1}{\sqrt{N}}e^{i\hat{\Omega}\tau_1}J_{+}(\vec{\Delta
k}_1)\ket{gg...g}$.

At first let us recall how an ideal $\pi$ pulse treat the atoms, which it swaps $g$ to $s$ and vice versa. One can describe that as $e^{i\pi/2 J_{x}}=e^{i\pi/2 \sum_{j=1}^{N}\sigma_{x}^{(j)}\otimes \mathbbm{1}}$, where $e^{i\pi/2 J_{x}}=\otimes_{k=1}^{N} i\sigma_{x}^{(k)}$, see \cite{footnote1}.

After applying the $\pi$ pulse, the operator $e^{i\hat{\Omega}t}$ leads to a rephasing of the collective excitation because for each term in the collective state the previously non-excited atoms now acquire phases whereas the previously excited atom doesn't. Finally, the stored pulse will be retrieved by applying the read field, leading to the emission of the readout photon, see FIG.\ref{fig1}(c). The readout part of the process can be represented by the operator $J_{+}(\vec{\Delta k}_2)=\sum_{k=1}^{N} e^{i\vec{\Delta k}_2.\vec{X}_k} \sigma_{+}^{(k)}\otimes \mathbbm{1}$, that is based on the same discussion as for the read-in. Eventually, the whole process comprised of creation of collective excitation, dephasing, $\pi$ pulse, rephasing and the read pulse that leads to the final state can be described as $\ket{\psi_{f}(\vec{\Delta k}_1,\vec{\Delta k}_2)}=\frac{1}{\sqrt{N}}J_{+}(\vec{\Delta k}_2)e^{i{\hat \Omega}\tau_2} e^{i\pi/2 J_{x}} e^{i{\hat \Omega}\tau_1} J_{+}(\vec{\Delta k}_1) \ket{gg...g}$.

Using Eq. (\ref{Hint}) and the analogous Hamiltonian for the readout process it is straightforward to include the quantum states of the light field for the Stokes and readout photons into the description. One sees that the emission amplitude for the readout photon can be obtained directly from the norm of the atomic state $\ket{\psi_{f}(\vec{\Delta k}_1,\vec{\Delta k}_2)}$. Note that there is no preferred direction for the Stokes photon emission \cite{stokes}. As before, we are not really interested in the absolute emission probability, but in how the probability varies as a function of the direction of emission for the read photon, and under the influence of errors in the control pulses. The collective enhancement happens  again under the phase-matching condition $\vec{\Delta k}_1+\vec{\Delta k}_2=0$.

In the ideal case, considering these conditions one can
easily derive that $\langle \psi_f(\vec{\Delta
k}_1,\vec{\Delta k}_2)| \psi_f(\vec{\Delta k}_1,\vec{\Delta
k}_2) \rangle = N$. The later result is based on
considering an ideal $\pi$ pulse in the calculations and it
takes place under the phase-matching condition,
$\vec{\Delta k}_1+\vec{\Delta k}_2=0$. As we studied in the
semi-classical approach, the re-emission in other
directions ($\vec{\Delta k}_1+\vec{\Delta k}_2\neq 0$) is
negligible for the case with large number of atoms in the
ensemble.

Our discussion here was focused on the case of the DLCZ
protocol, but the evolution of the atomic state is
identical in the other quantum memory protocols mentioned
in the introduction. The only difference is that in those
protocols the creation of the initial atomic excitation is
accompanied by the absorption of a single photon rather
than its emission.

\subsection{Global error in the $\pi$ pulse}

We now study the effect of imperfect $\pi$ pulse as one of the sources of error in the spin echo memories. The most general error in the $\pi$ pulse can be considered as a small rotation $\epsilon_k$ around a random direction $\hat{n}_k$ for each atom that can be represented by the operator $e^{\sum_{k=1}^{N}i\epsilon_{k}/2 \vec{\sigma}^{(k)}.\hat{n}^{(k)}\otimes\mathbbm{1}}$. This operator can show the effect of the inhomogeneity in the intensity of the $\pi$ pulse across the sample \cite{beavan}. For simplicity, we first consider a global error that can be interpreted as lack of accuracy in pulse shaping. The most general error that affects each atom differently is studied in the appendix. The corresponding operator for such error which is equally  distributed over all atoms is $e^{i\epsilon/2J.{\hat n}}=e^{i\epsilon/2\sum_{k=1}^{N} \vec{\sigma}^{(k)}.{\hat n}\otimes\mathbbm{1}}$.

Essentially, the question of studying the effect of the error in the $\pi$ pulse is reduced to the problem of analyzing norm of the state
\begin{eqnarray}\label{psi-final}
\ket{\psi_f(\vec{\Delta k}_1,\vec{\Delta k}_2,\epsilon)}&=&\frac{1}{\sqrt{N}} J_{+}(\vec{\Delta k}_2)e^{i{\hat \Omega}\tau_2} e^{i\pi/2 J_{x}}e^{i\epsilon/2J.{\hat n}} \\ \nonumber  &e&^{i{\hat \Omega}\tau_1} J_{+}(\vec{\Delta k}_1) \ket{gg...g}.
\end{eqnarray}

 In order to facilitate analyzing the norm of the state $\ket{\psi_f(\vec{\Delta k}_1,\vec{\Delta k}_2,\epsilon)}$ we now simplify the final state. Since we are only interested in its norm, any unitary operator can be used to simplify the state. By applying the unitary operator $e^{-i\pi/2 J_x}$ from the left and adding the identity $e^{i\pi/2 J_x}e^{-i\pi/2 J_x}$, one can represent the final state up to a unitary and a global phase as,
\begin{equation} \frac{1}{\sqrt{N}} J_{-}(\vec{\Delta k}_2) e^{-i{\hat \Omega}\tau_2} e^{i\epsilon/2J.{\hat n}} e^{i{\hat \Omega}\tau_1}J_{+}(\vec{\Delta k}_1) \ket{gg...g},
\end{equation}
where $J_{-}(\vec{\Delta k}_2)=\sum_{k=1}^{N} e^{i\vec{\Delta k}_2.\vec{X}_k} \sigma_{-}^{(k)}\otimes \mathbbm{1}$. Under the condition $\tau=\tau_1=\tau_2$ and by conducting some algebra one can derive that $ e^{-i{\hat \Omega}\tau} e^{i\epsilon/2J.{\hat n}} e^{i{\hat \Omega}\tau}= e^{\sum_{k=1}^{N}i\epsilon_{k}/2 \vec{\sigma}^{(k)}.\hat{n}'^{(k)}\otimes\mathbbm{1}}$, which shows rotation in a new direction $\hat{n}'^{(k)}= (n_x\cos{\Delta_k \tau} + n_y\sin{\Delta_k \tau})\hat{x}+(-n_x\sin{\Delta_k \tau} + n_y\cos{\Delta_k \tau})\hat{y}+n_z\hat{z}$. Finally, by applying the unitary $e^{\sum_{k=1}^{N}-i\epsilon_{k}/2 \vec{\sigma}^{(k)}.\hat{n}'^{(k)}\otimes\mathbbm{1}}$ the final state can be simplified to $\hat{\cal{O}} J_{+} \ket{gg...g}$ where the operator $\hat{\cal{O}}$ is given by,
\begin{eqnarray}
\hat{\cal{O}} = \frac{1}{\sqrt{N}}\sum_{k=1}^{N} e^{i\vec{\Delta k}_2.X_k} (\alpha\sigma_{-}^{(k)} + \beta e^{i\Delta_{k}\tau} \sigma_{z}^{(k)}
 + \gamma  e^{2i\Delta_{k}\tau} \sigma_{+}^{(k)})\otimes \mathbbm{1},
\end{eqnarray}
where $\alpha = \cos^{2}{(\epsilon/2)}+2i\sin{(\epsilon/2)}\cos{(\epsilon/2)} - {n_{z}}^{2}\sin^{2}{(\epsilon/2)}$, $\beta= -i\sin{(\epsilon/2)}\cos{(\epsilon/2)}(n_{x}-in_{y}) + n_{z}\sin^{2}{(\epsilon/2)}(n_{x}-in_{y})$ and $\gamma= \sin^{2}{(\epsilon/2)}(n_{x}-in_{y})^{2}$.
This simplification allows us to distinguish three different terms in the final state based on the effect of $\hat{\cal{O}}J_{+}$ on the $\ket{gg...g}$. The first term corresponds to $\sigma_{-}^{(k)}\sigma_{+}^{(j)}\ket{gg...g}$ which gives $\delta_{jk}\ket{gg...g}$. The other combination $\sigma_{z}^{(k)}\sigma_{+}^{(j)}$ in $\hat{\cal{O}}J_{+}$ yields $(-1)^{\delta_{jk}}\ket{g..s^{(j)}..g}$.The later gives $N^2$ terms that contains one excitation. Finally, the $\sigma_{+}^{(k)}\sigma_{+}^{(j)}$ leads to $N(N-1)$ terms with two excitations. One can benefit from these terms to represent the final state as
\begin{eqnarray}\label{psif}\alpha \ket{\psi_1}
+ \beta \ket{\psi_2}
 + \gamma \ket{\psi_3},
\end{eqnarray}
where $\ket{\psi_1} = \frac{1}{\sqrt{N}}\sum_{j=1}^{N}
e^{i(\vec{\Delta  k}_{1}+\vec{\Delta  k}_{2}).\vec{X}_{j}}
\ket{g..g}$, $\ket{\psi_2}=\frac{1}{\sqrt{N}}\sum_{j,k=1}^{N} (-1)^{\delta_{jk}} e^{i\Delta_{k}\tau}
e^{i\vec{\Delta  k}_{1}.\vec{X}_{j}}
e^{i\vec{\Delta  k}_{2}.\vec{X}_{k}}
\ket{g..s^{(j)}..g}$ and $\ket{\psi_3}=\frac{1}{\sqrt{N}}\sum_{j,k=1, j\neq k}^{N}  e^{2i\Delta_{k}\tau}
e^{i\vec{\Delta  k}_{1}.\vec{X}_{j}}
e^{i\vec{\Delta  k}_{2}.\vec{X}_{k}}
\mid g..s^{(j)}..s^{(k)}..g\rangle$. Obviously, because of different number of excitations, these terms are perpendicular. Thus, one can study the norm of the final state, easily.

As it can be seen from the eq. (\ref{psif}), because of $\delta_{jk}$, the first term corresponds to the directional emission of the readout photon. As we discussed in the previous section, see FIG. (\ref{fig2}), the readout is strongly peaked around the direction for which all of the single-atom re-emissions can constructively interfere with each other. The readout intensity at the other directions (the non-directional background) is suppressed by the ratio $1/N$ that is a quite small number for a typical $N\sim 10^7-10^9$. Hence, the $\alpha$ term determines the intensity of the echo. In other words, shining the read laser with $\vec{k}_{r}=-\vec{k}_{w}$ result in the emission of a readout photon that its intensity is peaked around $\vec{k}_{ro}=-\vec{k}_{s}$. The non-directional noise in the re-emission can be attributed to the terms that correspond to $\beta$ and $\gamma$ coefficients, because taking average over position of the atoms leads to randomly distributed re-emission from each single atom. It can be shown that the norm of the state in Eq. (\ref{psif}) is
\begin{eqnarray}\label{norm}
&\frac{ |  \alpha |^{2}}{N}&| \sum_{j=1}^{N} e^{i(\vec{\Delta  k}_{1}+\vec{\Delta  k}_{2}) .\vec{X}_{j}} |^{2}
 \\ \nonumber  + &\frac{| \beta |^{2}}{N}& \sum_{j=1}^{N}|\sum_{k=1}^{N}(-1)^{\delta_{jk}}e^{i\Delta_{k}\tau} e^{i(\vec{\Delta  k}_{1}).\vec{X}_{j}}
e^{i(\vec{\Delta  k}_{2}).\vec{X}_{k}} |^{2} + \frac{1}{N}\sum_{j,k=1, j\neq k}^{N} | \gamma|^{2}.
\end{eqnarray}

{\it Efficiency reduction.} In order to study the worst case scenario, upper bound of the noise strength and lower bound of the echo intensity have to be studied separately. Considering the phase-matching condition it can be shown that $I_{echo}\propto N |\alpha |^{2}$. For small errors, $\epsilon\ll 1$ and keeping terms up to ${\cal{O}}(\epsilon^{3})$, it can be shown that the following lower bound can be achieved for $n_z=0$,
\begin{equation}\label{echoint2}
| \alpha|^{2}\leq (1-2(\epsilon/2)^2).
\end{equation}
As in the ideal case the signal intensity is proportional
to $1$, one can analyze the efficiency reduction. So from
quantum mechanical point of view the efficiency reduction
factor is given by $1-2(\epsilon/2)^2$ that gives the same
results for the typical 1\% error as the semi-classical
calculation.

{\it Noise.} With the aim of studying the intensity of the noise that is proportional to the $\frac{1}{N}| \beta|^2 \sum_{j=1}^{N}|\sum_{k=1}^{N}(-1)^{\delta_{jk}}e^{i\Delta_{k}\tau} e^{i\vec{\Delta  k}_{1}.\vec{X}_{j}}
e^{i\vec{\Delta  k}_{2}.\vec{X}_{k}} |^{2}
+ \frac{1}{N}\sum_{j,k=1, j\neq k}^{N} | \gamma|^{2}$ in the limit of the small errors, $\epsilon\ll 1$, we need to discuss the effect of $e^{i\Delta_k \tau}$. For long enough times that $\tau$ is comparable with $\frac{1}{\Gamma}$, where $\Gamma$ is the inhomogeneous linewidth, the $e^{i\Delta_k \tau}$ becomes a completely random phase. It can be shown that for random phases $\phi_k$, $| \sum_{k} e^{i\phi_k}c_{k}|^2= |\sum_{k,l}c_{k}c_{l}^*e^{i(\phi_k-\phi_l)}|= \sum_{k}|c_{k}|^2$, because $\langle e^{i(\phi_k-\phi_l)} \rangle=\delta_{kl}$. Hence we can treat the term that corresponds to $\beta$ as $\frac{1}{N}|\beta|^2\sum_{j,k=1}^{N}| (-1)^{\delta_{jk}} e^{i\vec{\Delta  k}_{1}.\vec{X}_{j}}
e^{i\vec{\Delta  k}_{2}.\vec{X}_{k}}|^2=N |\beta|^2$. Taking these considerations into account shows that the intensity of the noise is proportional to
\begin{eqnarray}\label{noiseint1}
\frac{1}{N}|&\beta&|^2\sum_{j=1}^{N}|\sum_{k=1}^{N}(-1)^{\delta_{jk}}e^{i\Delta_{k}\tau} e^{i(\vec{\Delta  k}_{1}).\vec{X}_{j}}
e^{i(\vec{\Delta  k}_{2}).\vec{X}_{k}} |^{2} \\ \nonumber &+&\frac{N(N-1)}{N}|\gamma|^{2} \\ \nonumber &\leq& N (\epsilon/2)^2 \max(n_z^2(n_x^2+n_y^2)(\epsilon/2)^2+ n_x^2+n_y^2)\\ \nonumber &+& \frac{N^2-N}{N}(\epsilon/2)^4\max{|(n_{x}-in_{y})^{2}|^2}\\ \nonumber
&\approx& N (\epsilon/2)^2 + {\cal O}(\epsilon^{4}).
\end{eqnarray}
So the choice of uniformly directed error with $n_z=0$ gives the upper bound for the noise. Obviously, the second term in the noise intensity is proportional to $\epsilon^4$ that is negligible for the small errors. Then it leads to $I_{noise}\propto N (\epsilon/2)^2$. These results allow us to find the upper bound for the noise-to-signal ratio.

Let us recall the results for the noise and the echo from the semi-classical approach in eqs. (\ref{semiintecho},\ref{seminoiseint}). Obviously, the Taylor expansion of the results obtained in semi-classical treatment and eliminating the terms ${\cal O}(\epsilon^4)$ and higher, demonstrates the agreement between the results of the both approaches in the limit of small errors ($\epsilon\ll 1$).

In the quantum mechanical approach the noise is proportional to the term given in eq. (\ref{noiseint1}). The amplitude of the noise shows a correspondence with the fluorescent radiation in the semi-classical approach. Indeed the spatial phase dependence implies that the direction of the emission of the noise from each atom varies from one to another, leading to a non-directional noise. Thanks to collective enhancement, the non-directional noise will not swamp the echo signal, for realistic control pulse accuracy \cite{beavan}.

\section{Imperfect initial state and $\pi$ pulse}\label{sec-imperfect}
An imperfection in the $\pi$ pulse is not the only source of inefficiency in the spin echo memories. In our calculations so far we have assumed an ideal situation where all the atoms are initially in the ground state. Our quantum mechanical approach allows us to study the effect of an imperfect initial state with $n$ atoms excited to the state $s$. This can happen in experiments as a result of an imperfect optical pumping in the initialization of the atomic ensemble. Without loss of generality, we can consider the initial state $\ket{g..gs..s}$ that has $n$ sorted excited atoms instead of randomly positioned excited atoms. Applying the operator ${\cal O}J_+$ gives the final state. We expect the directional echo as result of applying $\sum_{j,k=1}^{N}\sigma_{-}^{(k)}\sigma_{+}^{(j)}$ on the initial state $\ket{g..gs..s}$. In contrast to the perfect initial case, this will lead to $N-n$ terms with $n$ excitations correspond to $\ket{g..gs..s}$, and also $(N-n)n$ terms connected with $\ket{g..s^{(j)}..gs..g^{(k)}..s}$ which has $n$ excitations. By conducting some algebra, one can easily show that only the first case gives $\delta_{jk}$ which leads to directional (collectively enhanced) re-emission under the phase-matching condition. This implies that by considering imperfection in both the initial state and the $\pi$ pulse, intensity of the echo is proportional to
\begin{equation}\label{impinitial}
I_{echo}\propto \frac{(N-n)^2}{N}(1-2(\epsilon/2)^2).
\end{equation}
The imperfection in the initial state reduces the intensity of the echo and also introduces a new source of the non-directional noise. Previously, we analyzed the noise that corresponds to $\ket{\psi_2}$ and $\ket{\psi_3}$ by finding the upper bound for $\beta$ and $\gamma$. Now, the $(N-n)n$ terms of the form $\ket{g..s^{(j)}..gs..g^{(k)}..s}$ also contribute to the noise.
 Consequently,  as the upper bound, for the term that corresponds to $\alpha$ can be achieved for $n_z=1$, one can obtain the following upper bound for the noise-to-signal ratio $r$,
\begin{equation}\label{impratio}
r\leq \frac{(N-n)n+N^2(\epsilon/2)^2}{(N-n)^2(1-2(\epsilon/2)^2)}.
\end{equation}
Investigating this equation it can be seen that the two
different sources of error compete in producing noise. For
the case with small fraction of excitations in the initial
state, $\frac{n}{N}\ll 1$,  such that
$\epsilon^2>\frac{4n}{N}$ then the error in $\pi$ pulse is
the dominant term in the noise. Otherwise, in case of
$\frac{4n}{N}>\epsilon^2$ the imperfection in the initial
state plays an important role in increasing the upper bound
of the noise-to-signal ratio. For instance, $2\%$ of the
atoms not in the ground state, which is a typical value
\cite{hedges09}, will cause a $2\%$ error. The error in the
$\pi$ pulse would have to be as large as $\epsilon=0.4$ in
order to be comparable in importance.

\section{Semi-classical treatment of multiple $\pi$ pulses}

So far we have studied the application of a single $\pi$ pulse. Much longer storage times are achievable by applying sequences of $\pi$ pulses. This is also known as bang-bang control. We analyze this case using the semi-classical approach. We have seen before that semi-classical and quantum treatment lead to identical conclusions for small errors. The intensity of the signal after applying $m$ pairs of $\pi$ pulses is given by,
\begin{equation}\label{gen1}
I_{echo}\approx I_{0} N^2 \xi^2 (1-(2m^2-m+1)\epsilon^2/2).
\end{equation}
The intensity that is associated to the noise also can be approximated by $I_{noise}\approx I_{0}N m^2\epsilon^2/2$. Thus, after applying sequence of the rephasing pulses, the noise-to-signal ratio reads
\begin{equation}\label{gen2}
r\approx m^2\epsilon^2/2,
\end{equation}
for $m\epsilon\ll 1$. For instance, with typical 1\% error
in the $\pi$ pulse one can benefit from 30 pairs of
rephasing pulses, while still achieving an efficiency
factor of 91\% and a noise-to-signal ratio of 0.04. If the
pulses are 4 msec apart as in Ref. \cite{longdell05}, 30
pairs of pulses correspond to a storage time of 240 msec.
Ref. \cite{beavan} managed to increase the $T_2$ time to 1
sec, by proper alignment of the magnetic field, that is
much longer than 80 msec in Ref. \cite{longdell05}. This
field configuration would allow one to put the rephasing
pulses further apart. For example, one can consider the
rephasing pulses 40 msec apart, giving a 2.4 sec storage
time with an efficiency factor of 91\% and a
noise-to-signal ratio of 0.04.

\section{Discussion and Conclusion}

We have shown that realistic imperfections in the $\pi$
pulses only have small effects on the retrieval efficiency
of quantum memories, and that the corresponding noise is
also acceptable. The latter fact is due to collective
interference. While the emission of the read photon is
strongly enhanced in the direction given by the phase
matching condition, the noise due to pulse imperfections is
non-directional. As a consequence, even the use of great
numbers of $\pi$ pulses (bang-bang control) is realistic at
the quantum level. This settles the question raised by Ref.
\cite{molmer04}.

We have also studied errors due to imperfect
optical pumping, i.e. imperfect initial state preparation.
We find that these errors are likely to dominate over $\pi$
pulse errors in many circumstances, but that they can also
be kept at an acceptable level.

Our discussion was phrased in terms of the DLCZ protocol
\cite{dlcz} for concreteness. However, the same results
apply to other ensemble-based quantum memory protocols
including those based on electromagnetically induced
transparency \cite{eit}, off-resonant Raman transitions
\cite{raman}, controlled reversible inhomogeneous
broadening \cite{crib}, and atomic frequency combs
\cite{afc}.

In conclusion, the prospects for the use of spin-echo
techniques in light-matter interfaces at the quantum level
are very good. We hope that our work will further encourage
experimental work in this direction.

{\it Acknowledgments.} This work was supported by an NSERC Discovery Grant, an AITF New Faculty Award, the EU project Q-essence, the Swiss NCCR Quantum Photonics, and iCORE. We thank M. Afzelius, N. Gisin and E. Saglamyurek for
useful discussions.

\appendix
\section{\label{appendix} Spatial inhomogeneity in the rf pulse}
As we discussed, the main error in rf pulses is the variance in the intensity of the rf pulse that leads to inhomogeneity of the rf pulse across the sample \cite{beavan}. Consequently, the error would be different for the atoms at the different positions. Here we extend our analysis to study such errors.

We study the spatial inhomogeneity in the rf pulse by considering $e^{\sum_{k=1}^{N}i\epsilon_{k}/2 \vec{\sigma}^{(k)}.\hat{n}^{(k)}\otimes\mathbbm{1}}$ as the operator which represents the error. Therefore, based on eq. (\ref{psi-final}) the final state is,
\begin{eqnarray}\label{app-psi-final}
\ket{\psi_f(\vec{\Delta k}_1,\vec{\Delta k}_2,\epsilon)} &=& \frac{1}{\sqrt{N}} J_{+}(\vec{\Delta k}_2)e^{i{\hat \Omega}\tau_2} e^{i\pi/2 J_{x}}\\ \nonumber &e&^{\sum_{k=1}^{N}i\epsilon_{k}/2 \vec{\sigma}^{(k)}.\hat{n}^{(k)}\otimes\mathbbm{1}}e^{i{\hat \Omega}\tau_1} J_{+}(\vec{\Delta k}_1) \ket{gg...g}.
\end{eqnarray}

We can follow the same approach as we used in the paper to simplify the final state that eases the rest of the calculation. Therefore, the final state $\ket{\psi_f(\vec{\Delta k}_1,\vec{\Delta k}_2,\epsilon)}$ up to a unitary operator and a global phase is
\begin{eqnarray}
\frac{1}{\sqrt{N}} J_{-}(\vec{\Delta k}_2) e^{-i{\hat \Omega}\tau_2} e^{\sum_{k=1}^{N}i\xi_{k} \vec{\sigma}^{(k)}.\hat{n}^{(k)}\otimes\mathbbm{1}} e^{i{\hat \Omega}\tau_1} J_{+}(\vec{\Delta k}_1) \ket{gg...g}.
\end{eqnarray}
Considering the $\pi$ pulse at the middle of the process, $\tau=\tau_1=\tau_2$, and by conducting some algebra one can derive that $ e^{-i{\hat \Omega}\tau} e^{\sum_{k=1}^{N}i\xi_{k} \vec{\sigma}^{(k)}.\hat{n}^{(k)}\otimes\mathbbm{1}} e^{i{\hat \Omega}\tau}= e^{\sum_{k=1}^{N}i\xi_{k} \vec{\sigma}^{(k)}.\hat{n}'^{(k)}\otimes\mathbbm{1}}$, which shows rotation in a new direction $\hat{n}'^{(k)}= (n_x^{(k)}\cos{\Delta_k \tau} + n_y^{(k)}\sin{\Delta_k \tau})\hat{x}+(-n_x^{(k)}\sin{\Delta_k \tau} + n_y^{(k)}\cos{\Delta_k \tau})\hat{y}+n_z^{(k)}\hat{z}$. Finally, by applying the unitary $e^{\sum_{k=1}^{N}i\xi_{k} \vec{\sigma}^{(k)}.\hat{n}'^{(k)}\otimes\mathbbm{1}}$ the final state can be simplified to
\begin{equation}\label{app-psif1}
\ket{\psi_f(\vec{\Delta k}_1,\vec{\Delta k}_2,\epsilon)} = \hat{\cal{O}} J_{+} \ket{gg...g},
\end{equation}
where the operator $\hat{\cal{O}}$, which acts on $k^{th}$ atom is given by,
\begin{eqnarray}
\hat{\cal{O}} &=& \frac{1}{\sqrt{N}}\sum_{k=1}^{N} e^{i\vec{\Delta k}_2.\vec{X}_k} (\alpha^{(k)}\sigma_{-}^{(k)}
+ \beta^{(k)} e^{i\Delta_{k}\tau} \sigma_{z}^{(k)}\\ \nonumber
&+& \gamma^{(k)}  e^{2i\Delta_{k}\tau} \sigma_{+}^{(k)})\otimes \mathbbm{1},
\end{eqnarray}
and $\alpha^{(k)} = \cos^{2}{(\epsilon_{k}/2)}+2i\sin{(\epsilon_{k}/2)}\cos{(\epsilon_{k}/2)} - {n_{z}^{(k)}}^{2}\sin^{2} {(\epsilon_{k}/2)}$, $\beta^{(k)}= -i\sin{(\epsilon_{k}/2)}\cos{(\epsilon_{k}/2)}(n_{x}^{(k)}-in_{y}^{(k)}) + n_{z}^{(k)}\sin^{2}{(\epsilon_{k}/2)}(n_{x}^{(k)}-in_{y}^{(k)})$ and $\gamma^{(k)}= \sin^{2}{(\epsilon_{k}/2)}(n_{x}^{(k)}-in_{y}^{(k)})^{2}$.
Finally, the state can be simplified to
\begin{eqnarray}\label{app-psif}
&|&\psi_f(\vec{\Delta k}_1,\vec{\Delta k}_2,\epsilon)\rangle= \frac{1}{\sqrt{N}}\sum_{j=1}^{N} \alpha^{(j)}
e^{i(\vec{\Delta  k}_{1}+\vec{\Delta  k}_{2}).\vec{X}_{j}}
\ket{g..g} \\ \nonumber
&+& \frac{1}{\sqrt{N}}\sum_{j,k=1}^{N} (-1)^{\delta_{jk}}\beta^{(k)} e^{i\Delta_{k}\tau}
e^{i(\vec{\Delta  k}_{1}).\vec{X}_{j}}
e^{i(\vec{\Delta  k}_{2}).\vec{X}_{k}}
\ket{g..s^{(j)}..g} \\ \nonumber
 &+& \frac{1}{\sqrt{N}}(\sum_{j,k=1, j\neq k}^{N} \gamma^{(k)} e^{2i\Delta_{k}\tau}
e^{i(\vec{\Delta  k}_{1}).\vec{X}_{j}}
e^{i(\vec{\Delta  k}_{2}).\vec{X}_{k}})\\ \nonumber
&|& g..s^{(j)}..s^{(k)}..g\rangle. \nonumber
\end{eqnarray}
Fortunately, having different number of excitations in the final state simplifies the calculation of the norm.

As we discussed, the first term contributes in directional emission of the readout photon. Hence, the $\alpha^{(k)}$s play role in finding the intensity of the echo. In other words, shining the read laser with $\vec{k}_{r}=-\vec{k}_{w}$ result in the emission of a readout photon that is peaked around $\vec{k}_{ro}=-\vec{k}_{s}$. The non-directional noise in the re-emission can be attributed to the terms corresponding to $\beta^{(k)}$ and $\gamma^{(k)}$ coefficients, because average over the position of the atoms leads to randomly distributed re-emission. It can be shown that the following gives norm of the $\ket{\psi_f}$ in eq. (\ref{app-psif}),
\begin{eqnarray}\label{app-norm}
A^2 &=& \frac{1}{N}\mid\sum_{j=1}^{N}\alpha^{(j)} e^{i(\vec{\Delta  k}_{1}+\vec{\Delta  k}_{2}) .\vec{X}_{j}} \mid^{2}
 \\ \nonumber  &+& \frac{1}{N}\sum_{j=1}^{N}\mid\sum_{k=1}^{N}(-1)^{\delta_{jk}}e^{i\Delta_{k}\tau}\beta^{(k)} e^{i(\vec{\Delta  k}_{1}).\vec{X}_{j}}
e^{i(\vec{\Delta  k}_{2}).\vec{X}_{k}} \mid^{2} \\ \nonumber
&+& \frac{1}{N}\sum_{j,k=1, j\neq k}^{N} \mid \gamma^{(k)}\mid^{2}.
\end{eqnarray}

In order to analyze the efficiency reduction the lower bound of echo intensity have to be studied. Considering the phase-matching condition it can be shown that $I_{echo}\propto \frac{1}{N}|\sum_{j=1}^{N}\alpha^{(j)} |^{2}$. For small errors, $\epsilon_j\ll 1$ and keeping terms to ${\cal{O}}(\epsilon^{3})$, it can be shown that the following lower bound can be achieved for $n_z^{(j)}=0$
\begin{equation}\label{app-echoint2}
\frac{1}{N}|\sum_{j=1}^{N}\alpha^{(j)} |^{2}\leq N (1-2(\epsilon_{max}2/)^2),
\end{equation}
where $\epsilon_{max}$ is the largest error and $N$ is number of the atoms.
With the aim of studying the intensity of the noise that is proportional to the $\frac{1}{N}\sum_{j=1}^{N}\mid\sum_{k=1}^{N}(-1)^{\delta_{jk}}e^{i\Delta_{k}\tau}\beta^{(k)} e^{i(\vec{\Delta  k}_{1}).\vec{X}_{j}}
e^{i(\vec{\Delta  k}_{2}).\vec{X}_{k}} \mid^{2}
+ \frac{1}{N}\sum_{j,k=1, j\neq k}^{N} \mid \gamma^{(k)}\mid^{2}$ for the small errors, $\epsilon_j\ll 1$, one needs to consider the $e^{i\Delta_k \tau}$ as a random phase, that takes place for long enough times that $\tau$ is comparable with the $\frac{1}{\Gamma}$. It can be shown that for random $\phi_k$, $|\sum_{k} e^{i\phi_k}\beta^{(k)}|^2= |\sum_{k,l} \beta^{(k)}\beta^{(l)^*}e^{i(\phi_k-\phi_l)}|^2=\sum_{k}|\beta^{(k)}|^2$, because $\langle e^{i(\phi_k-\phi_l)}\rangle=\delta_{kl}$. Hence, one can conclude that $|\sum_{k=1}^{N}(-1)^{\delta_{jk}}e^{i\Delta_{k}\tau}\beta^{(k)} e^{i(\vec{\Delta  k}_{1}).\vec{X}_{j}}
e^{i(\vec{\Delta  k}_{2}).\vec{X}_{k}} |^{2}$ can be approximated as $\sum_{k=1}^{N}| (-1)^{\delta_{jk}}\beta^{(k)} e^{i(\vec{\Delta  k}_{1}).\vec{X}_{j}}
e^{i(\vec{\Delta  k}_{2}).\vec{X}_{k}}|^2=\sum_{k=1}^{N}| (-1)^{\delta_{jk}}\beta^{(k)}|^2$. By taking these considerations into account and keeping terms up to ${\cal O}(\epsilon^4)$ give the noise intensity upper bound as
\begin{eqnarray}\label{app-noiseint1}
I_{noise}&\leq&  N \max(n_z^{(k)^2}(n_x^{(k)^2}+n_y^{(k)^2})(\epsilon_k/2)^4  \\ \nonumber &+& (n_x^{(k)^2}+n_y^{(k)^2})(\epsilon_k/2)^2)\\ \nonumber
&+& \frac{N^2-N}{N}\max((\epsilon_k/2)^4|(n_x^{(k)}-in_y^{(k)})^2|^2) \\ \nonumber
&=& N(\epsilon_{max}/2)^2+{\cal O}(\epsilon^4),
\end{eqnarray}
which implies that a uniformly directed error with $n_z=0$ gives the upper bound for the noise. Obviously, the second term in the noise intensity is proportional to $\epsilon^4$ that is negligible for the small errors. Then it leads to $I_{noise}\leq N(\epsilon_{max}/2)^2$. This shows that the fully quantum mechanical treatment for an inhomogeneous error is in good agreement with the results for global errors derived in the paper.


\begin{thebibliography}{99}
\bibitem{footnote1} All of the operators in the power are local and they commute so $e^{i\pi \sum_{j=1}^{N}\sigma_{x}^{(j)}\otimes \mathbbm{1}}=\prod_{j=1}^{N} e^{i\pi \sigma_{x}^{(j)}\otimes \mathbbm{1}}$. It can be shown that $e^{A\otimes\mathbbm{1}}=e^{A}\otimes\mathbbm{1}$ and $e^{\mathbbm{1}\otimes B}=\mathbbm{1}\otimes e^{B}$ for any operator $A$ and $B$.
\end{thebibliography}
\end{document}